\documentclass{article}
\font\Bbb = msbm10.tmf
\def\const {{\hbox{const}}}
\def\R {{\hbox {\Bbb R}}}
\def\O {{\cal O}}
\def\o {\hbox{o}}

\newtheorem{Lemma}{Lemma}[section]
\newtheorem{Th}{Theorem}[section]
\newtheorem{cons}{Corollare}[section]

\title
{\bf Liouville equation under perturbation}
\author
{\bf L.A. Kalyakin
\\
Institute of Mathematics,  Ufa Sci. Centre, of  Russian Acad. of
Sci.\\ Institute of Mathematics, 112,Chernyshevsky str., Ufa,
\\450000, Russia\\
E-mail: klen@imat.rb.ru
\thanks
{This research has been supported by the Russian
Foundation of the Fundamental Research under Grants
99-01-00139, 96-15-96241} }
\date{December, 16, 1999}
\begin{document}
\maketitle

\begin{abstract}
Small perturbation of the Liouville equation under smooth
initial data is considered. Asymptotic solution which is
available for a long time interval is constructed by the
two scale method.
\end{abstract}

The Cauchy problem for the Liouville equation with a
small perturbation
$$
\partial_t^2u   -\partial_x^2u
+8\exp u =\varepsilon{\bf F}[u],
\quad 0< \varepsilon \ll 1,
\eqno (0.1)
$$
$$
u |_{t=0}=\psi_0(x), \
\partial_tu|_{t=0}=\psi_1(x),\quad x\in\R
\eqno (0.2)
$$
is considered. The problem is not suit under soliton
perturbation theory because the Liouville equation has
no any soliton solution. Initial functions are here
arbitrary, ones are smooth and decay rapidly at
infinity
\ ${\psi}_0,{\psi}_1(x)=\O(x^{-N}),
\ |x|\to\infty,\ \forall\, N$. So we deal with a
smooth solution; the case of singular solutions was
considered in [1].

The perturbation operator is determined by two smooth
functions $F_1,F_2$:
$$
{\bf F}[u]=\partial_xF_1(\partial_xu,\partial_tu)+
\partial_tF_2(\partial_xu,\partial_tu).
\eqno (0.3)
$$
The purpose is to construct an asymptotic approach of the
solution $u(x,t;\varepsilon)$ as $\varepsilon\to 0$
uniformly over long time interval $\{ x\in\R,\ 0<t\leq
\O(\varepsilon^{-1})\}$.

{\bf Results.} {\it 1. The solution of the unperturbed
problem (as $\varepsilon=0$) decomposes asymptotically
at infinity (as $t\to\infty$) on two simple waves
which travel on a decreasing background\
$u(x,t;0)=-4t+A^0_\pm(s^\pm)+\O((s^\mp)^{-N}),\
s^\mp\to\mp\infty$. 2. The structure of the asymptotic
solution as $\varepsilon\to 0$ remains the same
$u(x,t;\varepsilon)\approx
-4t+A_\pm(s^\pm,\tau)+\O(\varepsilon)$
for long times $t\approx\varepsilon^{-1}$. The
perturbation affects only a slow deformation of the
waves $A_\pm=A_\pm(s^\pm,\tau)$ on the slow time scale
$\tau=\varepsilon t$. 3. The deformation of the waves
is described by the first order PDE's
$$
\pm 2\partial_\tau A_\pm=H_\pm(\partial_s A_\pm),\quad
s\in\R, \ \tau >0,
\eqno (0.4)
$$
where
$
H_\pm(B)=F_1(B,-4\pm B)- F_1(0,-4)\pm  F_2(B,-4\pm
B)\mp F_2(0,-4).
$
The initial data are here taken from the fast time
asymptotics of the unperturbed solution
$$
A_\pm(s,0)=A^0_\pm(s)
\eqno (0.5)
$$
}

This result is very close to the case of linear wave
under perturbation, which slow deformation is described
by the Hopf equation, [2].

We use here the two scale method and construct an
asymptotic solution as a piece of the asymptotic series
$$
u\approx\sum_{n=0}^\infty\varepsilon^n\stackrel{n}{u}
(x,t,\tau),\ \tau=\varepsilon t,
\quad \varepsilon\to 0.
\eqno (0.6)
$$

{\bf Remark.} The direct asymptotic expansion
$
u\approx\sum\varepsilon^n\stackrel{n}{u}(x,t)
$
does not provide approach to the solution over long
time interval $t\approx\O(\varepsilon^{-1})$ through
the secular terms in corrections.

\section
{Unperturbed problem ($\varepsilon=0$)}

General solution of the nonlinear equation
$$
\partial_t^2\phi   -\partial_x^2\phi
+8\exp\phi  =0,
\ (x,t)\in R^2
$$
is given (Liouville, [3]) by the formula
$$
\phi (x,t)=\ln {{r_+^{\prime}(s^+)r_-^{\prime}(s^-)}\over
{r^2(s^+,s^-)}},\quad  r=r_+(s^+)+r_-(s^-), \
s^{\pm}=x\pm t
\eqno (1.1)
$$
with arbitrary functions $r_\pm$. Initial conditions
(0.2) give two ODE's for the $r_\pm$ which may be
linearized by change of variable
$r_\pm^\prime=w/\rho_\pm^2,\ (\forall\, w=\const\neq
0)$, so that
$$
\rho_\pm^{\prime\prime}-\Psi_\pm (x)\rho_\pm =0.
\eqno (1.2)
$$
The potentials $\Psi_\pm (x)$ are determined by the
original initial data, [4]
$$
\Psi_\pm(x)=\exp(\psi_0)+
\Big({{\psi_0^\prime \pm\psi_1}\over 4}\Big)^2-{1\over
4}(\psi_0^\prime
\pm\psi_1)
^\prime.
$$
So the Cauchy problem for the Liouville equation is
integrable.

\begin{Lemma}
Let as
$
\psi_0,\psi_1(x)=\O(x^{-N}),\ |x|\to\infty,\ \forall\, N,
$
and equations (1.2) are not on spectrum. Than the
solution of the Cauchy problem for the Liouville equation
has an asymptotics
$$
\phi (x,t)=-4t+ A_\pm (s^\pm )
+\O((s^\mp )^{-N})+\O(e^{-4t}),\ s^\mp\to\mp\infty,\
t\to\infty
$$
with the matching property
$$
A_\pm (s)=\cases{-\ln a^2 +\O(s^{-N}),\
s\to\pm\infty,\ (a=\const\neq 0),
\cr \O(s^{-N}),\ s\to\mp\infty .}
\eqno (1.3)
$$
\end{Lemma}

Functions $A_\pm (s)$ are reading from $r_\pm (s)$. If
the $r_\pm$ are fixed by the conditions at infinity
$$
r_\pm(x)=(1/a)\exp(2x)[1+\O(x^{-N})], \quad x\to
-\infty,
\eqno (1.4)
$$
$$
r_\pm (x)=a\exp(2x)[1+\O(x^{-N})], \quad x\to +\infty
\eqno (1.5)
$$
then
$$ A_+(s^+)=\ln\Big[{{\exp(2s^+)r_+^{\prime}(s^+)
}/{ar^2_+(s^+)}}\Big],\quad
A_-(s^-)=\ln\Big[{{\exp(-2s^-)r_-^{\prime}(s^-)
}/{a}}\Big].
\eqno (1.6)
$$
Functions  $A_\pm (s)$ do not depend on the choice of
$r_\pm(s)$ within  the Bianchi transform [5] and may be
used for parametrization of the general solution.

\begin{cons}
General solution of  the Liouville equation can be
parametrized by the pare of functions $A_\pm(s)$ which
have the matching property (1.3) so that
$$
\phi(x,t)=\Phi[A_+,A_-]\equiv
\ln {{r_+^{\prime}r_-^{\prime}}\over {(r_++r_-)^2}}
\eqno (1.7)
$$
in view of (1.6).
\end{cons}

\bigskip

\section
{ Linearized problem for the correction}

Corrections $\stackrel{n}{u}\ (n \geq 1)$ are obtained
from linear equations with corresponding initial
conditions
$$
\partial_t^2\stackrel{n}{u}
-\partial_x^2\stackrel{n}{u}+
8{{r_+^{\prime}r_-^{\prime}}\over
{r^2}}\stackrel{n}{u}=\stackrel{n}{f}(x,t;\varepsilon  ),
\quad
\stackrel{n}{u}|_{t=0}=\stackrel{n}{\psi}_0(x), \
\partial_t\stackrel{n}{u}|_{t=0}=\stackrel{n}{\psi}_1(x).
\eqno (2.1)
$$
The right sides are here determined by the previous
approaches. Dependence on the fast variables $x,t$ is
only determined from these equations.

General solution of the homogeneous linear equation is
given by the formula
$$
u_0(x,t;\varepsilon
)={{j_+^{\prime}}\over{r_+^{\prime}}}+
{{j_-^{\prime}}\over{r_-^{\prime}}}- 2{{j_++j_-}\over{r}}
$$
where $j_\pm=\stackrel{n}{j}_\pm(s^\pm)$ are arbitrary
functions. In context of the Cauchy problem they are
determined by the initial data. In this way pare linear
ODE's are obtained which can be solved in explicit form.

A similar formula with $j_\pm(s^\pm,t)$ may be used to
solve the nonhomogeneous linear equation (2.1). The
functions  $j_\pm(x,t)$ are defined from ODE's as well,
so that the solution is represented by the integral
$$
u(x,t)=\int_{s^-}^{s^+}\int_{s^-}^{\sigma^+}
K({s^+},{s^-},\sigma^+,\sigma^-)f(\sigma^+,\sigma^-)
\,d\sigma^- \,d\sigma^+
$$
taken over the characteristic triangle. The kernel $K$
is here expressed by means of $r_\pm$
$$
K({s^+},{s^-},\sigma^+,\sigma^-)=
{1\over{2r(s_+,s_-)r(\sigma_+,\sigma_-)}}
$$
$$
\Big\{
r_+(s^+)r_-(s^-) + r_+(\sigma^+)r_-(\sigma^-)
 +{1\over 2}\Big[r_+(s^+)-r_-(s^-)\Big]
\Big[r_+(\sigma^+)-r_-(\sigma^-)\Big]\Big\}.
$$

\begin{Lemma}
Let both the right side and the initial functions
decay rapidly at infinity. Than the solution of the
Cauchy problem for linearized equation is bounded and
has an asymptotics
$$
u(x,t)=U_\pm(s^\pm)+
\O((s^\mp)^{-N}),\ s^\mp\to\mp\infty,
$$
$$ U_\pm(s)=\cases{U+\O(s^{-N}),\ s\to\pm\infty,\
(U=\const),
\cr \O(s^{-N}),\ s\to\mp\infty .}
$$
\end{Lemma}

\section
{ Perturbed problem ($\varepsilon\neq 0$)}

We construct a formal asymptotic solution in the form
(0.6) where the leading order term is taken as a
solution of the unperturbed equation. The original
idea is to use $A_\pm$ - parametrization of this
solution
$
\stackrel{0}{u}=\Phi[A_+,A_-]\
$
as  it was pointed in (1.7). The second idea becomes from
the two scale method. It is assumed the functions
$A_\pm(s^\pm,\tau)$ depend on both fast $s^\pm=x\pm t$
and slow $\tau=\varepsilon t$ variables. The initial
values for the $A_\pm(s^\pm,\tau)$ as $\tau=0$ are taken
in (0.5) from the unperturbed solution. Ones are
calculated per the initial function $\psi_0,\psi_1(x)$
from the equations (1.2),(1.4),(1.6).

Dependence on the slow variable as $\tau>0$ is determined
by the differential equations obtained from the secular
condition which means the first order correction is
small:\ $\varepsilon\stackrel{1}{u}=\o(1),\
\varepsilon\to 0,\ \tau=\varepsilon t$\ uniformly over long time interval
$0<t\leq\O(\varepsilon^{-1})$.

The right side of the first order equation is given by
formula
$$
\stackrel{1}{f}(x,t,\tau)={\bf F}[\stackrel{0}{u}]
-2\partial_\tau\partial_t\stackrel{0}{u}.
$$
One can see from the lemma 2.1 that the secular condition
can be formulated through the right side as follows: The
right side tends to zero at infinity as $s^\pm\to 0$.
This requirement gives two equations
$$
\pm 2\partial_\tau\partial_sA_\pm={\bf F}[-4t+A_\pm].
$$
If we integrate these relations taking into account
boundary conditions
$$
A_\pm(s^\pm,\tau)\to 0 \quad as \quad s^\pm\to\mp\infty
$$
than the first order PDE's (0.4) are obtained.

\begin{Lemma}
1. If the perturbation operator has the form (0.3)
than the Cauchy problem for deformation equations
(0.4),(0.5) has the unique smooth solution on some
finite interval $0\leq\tau\leq\tau_0$. 2. Under such
functions $A_\pm(s^\pm,\tau)$ the secular condition is
hold for the first correction, i.e.
$\stackrel{1}{u}(x,t,\tau)$ is bounded uniformly for
all $x,t\in \R^2,\
\tau\in [0,\tau_0]$.
\end{Lemma}

\begin{cons}
The function $U_1(x,t,\varepsilon)=\stackrel{0}{u}
(x,t,\varepsilon
t)+\varepsilon\stackrel{1}{u}(x,t,\varepsilon t)$\ under
substitution in the equations (0.1),(0.2)  gives a
remainder of order $\O(\varepsilon^2)$ uniformly for all
$x\in\R,\ 0\leq t\leq\tau_0\varepsilon^{-1}$.
\end{cons}

\begin{Th}
Let the perturbation operator has the form (0.3) and
the initial functions in (0.2) are such that equations
(1.2) are not on spectrum (i.e. in general position).
Than the leading order term of the formal asymptotic
solution for the perturbed problem (0.1),(0.2) is
given by the Liouville formula (1.1) in which the
functions $r_\pm(s^\pm,\tau)$ depend on the additional
slow time $\tau=\varepsilon t$. Slow deformation of
the $r_\pm(s^\pm,\tau)$  is determined from equations
(0.4),(0.5),(1.6),(1.4).
\end{Th}


\begin{thebibliography}{99}
\bibitem {1}
L.A. Kalyakin, Teoret. Matemat. Fisika 118, 3 (1999)
390-396 (in Russian).

\bibitem {2}
L.A. Kalyakin, Math. USSR Sbornik 52, 1 (1985) 91-114
Matem. Sbornik. 124, 1 (1984) 96-120 (in Russian).

\bibitem {3}
J. Liouville, Journ. math. pure et appl. 18 (1853)
71-74.

\bibitem {4}
G.P.Jeorjadze, A.K. Pogrebkov, M.C. Polivanov, Teoret.
Matemat. Fisika 40, 2 (1979) 221-234 (in Russian).

\bibitem {5}
L. Bianchi, Ann. Sci. Norm. Super. Piza, Ser 1, 2 (1879)
26.


\end{thebibliography}
\end{document}